\documentclass{emulateapj}
\usepackage{apjfonts}
\def\Meszaros{M\'esz\'aros~}
\begin{document}

\title{Synchrotron self-Compton emission from  external shocks as the origin of the sub-TeV emission in GRB~180720B and GRB~190114C }

\author{Xiang-Yu Wang \altaffilmark{1,4}, Ruo-Yu Liu\altaffilmark{2}, Hai-Ming Zhang\altaffilmark{1,4}, Shao-Qiang Xi\altaffilmark{1,4} and Bing Zhang\altaffilmark{3}}

\altaffiltext{1}{School of Astronomy and Space Science, Nanjing University, Nanjing 210023, China; xywang@nju.edu.cn}

\altaffiltext{2}{Deutsches Elektronen Synchrotron (DESY), Platanenallee 6, D-15738 Zeuthen, Germany; ruoyu.liu@desy.de}
\altaffiltext{3}{Department of Physics, University of Nevada, Las Vegas, 4505 Maryland Parkway, Las Vegas, NV 89154-4002; zhang@physics.unlv.edu}
\altaffiltext{4}{Key laboratory of Modern Astronomy and Astrophysics (Nanjing University), Ministry of Education, Nanjing 210023, China}

\begin{abstract}
Recently, very high-energy photons above 100 GeV were reported to be detected from GRB~190114C and GRB~180720B  at, respectively, 100-1000 s and 10 hours after the burst. We model the available broad-band data of both GRBs with the synchrotron plus  synchrotron self-Compton (SSC) emission of the afterglow shocks.   We find that the sub-TeV emission of  GRB~180720B  can be interpreted as the  SSC emission from afterglow shocks expanding in a constant density circum-burst medium.  The SSC emission of GRB~190114C dominates over the synchrotron component   from GeV energies at $\sim100$~s, which can explain the possible hard spectrum of the GeV emission at this time. The extrapolated flux of this SSC component to sub-TeV energies can explain the high-significance detection of GRB~190114C by the MAGIC telescope.
The  parameter values (such as the circum-burst density and shock microphysical parameters) in the modeling are not unusual for both GRBs, implying that the detection of sub-TeV photons from these two bursts should be attributed to their large burst energies and low redshifts.
\end{abstract}


\section{Introduction}
Very high-energy (VHE) photons probe the most energetic particles accelerated in gamma-ray bursts (GRBs), so they are crucial to study the particle acceleration and radiation physics in GRBs.  Intense efforts have been made to detect VHE gamma-rays ($>100$ GeV) from GRBs (e.g., Aliu et al. 2014; Abramowski et al. 2014; Abeysekara et al. 2015), but it was only until recently that such VHE photons are detected from GRB~190114C and GRB~180720B (Mirzoyan 2019; Ruiz-Velasco 2019). MAGIC slewed to the direction of GRB 190114C about 50 s after the trigger and detected $>300$ GeV photons for the first 20 minutes from this burst with a significance of $>20\sigma$ (Mirzoyan 2019). HESS started to observe GRB~180720B at about 10 hr after the burst and detected $100-440$ GeV photons at such late times (Ruiz-Velasco 2019). Both GRBs have  relatively low redshifts, with $z=0.4245$  and $z=0.653$ for GRB~190114C and~GRB 180720B respectively (Selsing et al. 2019; Vreeswijk et al. 2019) . They are also bright bursts with the isotropic energies of $3\times10^{53}{\rm erg}$ and $6\times10^{53}{\rm erg}$, respectively (Hamburg et al. 2019; Frederiks et al. 2018).

It has been argued that high-energy photons above 100 MeV detected by Fermi Large Area Telescope (LAT) are produced by synchrotron radiation in the afterglow shocks (e.g., Kumar \& Barniol Duran 2009; Ghisellini et al. 2010;
Wang et al. 2010). However, the synchrotron emission has a maximum energy of $50 \Gamma \,{\rm MeV}$, where $\Gamma$ is the bulk Lorentz factor of the emitting region,  so it is hard to explain $>10$ GeV photons detected at $>100$ s where the shock has been decelerated, i.e., $\Gamma\le 200$ (Piran \& Nakar 2010). It was argued that these $>10$ GeV photons   should be produced by SSC emission in the afterglow shocks, supported by multi-band modeling of some LAT-detected GRBs (Wang et al. 2013). Indeed, afterglow SSC emission  has long predicted to be able to produce high-energy photons ({e.g., \Meszaros \& Rees 1993; Waxman 1997; Chiang \& Dermer 1999; Panaitescu \& Kumar 2000; Wang et al. 2001; Zhang \& \Meszaros 2001; Sari \& Esin 2001; Granot \& Guetta 2003; Fan \& Piran 2008; Beniamini et al. 2015}).

Recent detections of  sub-TeV emission from GRB~190114C and GRB~180720B strengthened the difficulty for the synchrotron radiation model.  One may naturally think about the inverse-Compton (IC) mechanism for such sub-TeV photons. Another interesting question is whether the sub-TeV emissions detected at quite different times from the two GRBs have a common origin.  In this paper, we will study whether the SSC mechanism can explain the sub-TeV emission of the two GRBs.
{Derishev \& Piran (2019) discussed the SSC mechanism for sub-TeV emission of GRB 190114C and explored the physical conditions in the emitting region of the afterglow of GRB 190114C. For simplicity, they considered a single-energy electron population. Here we perform modeling of the available broad-band data of both GRB 180720B and GRB 190114C using a realistic distribution (i.e., power-law distribution) for shock-accelerated electrons.}
In \S2, we first derive the light curves of SSC emission and compare them with the sub-TeV data of GRB~180720B. In \S3, we study whether the $\gamma\gamma$ absorption and Klein-Nishina (KN) suppression  affect the sub-TeV emission. In \S4, we model the observed light curves and spectral energy distribution (SED)  of the available multi-band data for both GRBs. Finally we give discussions and conclusions in \S5.

\section{The light curve of the SSC emission}
The temporal decay slope of SSC emission depends on the density profile of the circum-burst medium and the spectral regime of the observed frequency.  We first derive the slope and then compare it with the available data of sub-TeV emission.
As a rough approximation, the afterglow SSC spectrum can be described by broken power-laws with two break frequencies at $\nu_m^{\rm IC}$ and $\nu_c^{\rm IC}$ and a peak flux at $F_m^{\rm IC}$, generally resembling the spectrum of the synchrotron emission (Sari \& Esin 2001). For a stellar wind medium of $n\propto R^{-2}$, we have $\nu_m^{\rm IC} \propto t^{-2}$, $\nu_c^{\rm IC} \propto t^{2}$, and $F_m^{\rm IC}\propto \tau F_m^{\rm syn}$, where $\tau$ is the optical depth of the inverse-Compton (IC) scattering, which scales as $\tau\propto t^{-1/2}$. The frequency of VHE emission is typically  above $\nu_m^{\rm IC}$ from very early times. Thus, for the wind environment,
\begin{equation}
F_{\nu}= \left\{
\begin{array}{lll}
F_m^{\rm IC} \left(\frac{\nu}{\nu_m^{\rm IC}}\right)^{-\frac{p-1}{2}}\propto t^{-p}, \,\,\,\, \nu_m^{\rm IC}<\nu<\nu_c^{\rm IC}  \\
F_m^{\rm IC} (\frac{\nu}{\nu_c^{\rm IC}})^{-\frac{1}{2}}\propto t^0,  \,\,\,\, \nu_c^{\rm IC}<\nu<\nu_m^{\rm IC} \\
F_m^{\rm IC} ({\nu_m^{\rm IC}})^{\frac{p-1}{2}}({\nu_c^{\rm IC}})^{\frac{1}{2}} \nu^{-\frac{p}{2}} \propto t^{-(p-1)}, \,\,\,\, \nu>\max(\nu_m^{\rm IC}, \nu_c^{\rm IC})
\end{array}
\right.
\end{equation}
We note that the KN effect has not been taken into account in these scalings.
{The energy flux of GRB~180720B observed by HESS at 10~hr is about $5\times 10^{-11}{\rm erg cm^{-2} s^{-1}}$ in 100-440~GeV (Ruiz-Velasco 2019). Assuming a   mildly rising $\nu F_\nu$ spectrum, the  flux at $\sim1-10$ GeV would be at least $\ga 10^{-11}{\rm erg \, cm^{-2} s^{-1}}$. Since  GeV frequency is expected to be below $\nu_c^{\rm IC}$ (especially at later times since $\nu_c^{\rm IC}$ increases rapidly with time), the  GeV flux contributed by the SSC component should decay as $t^{-p}$. To be conservative, we assume a decay slope of $t^{-p}\sim t^{-2}$, then the SSC flux extrapolated  to $t=100 $ s would be nearly $\ga 2\times 10^{-6}{\rm erg \, cm^{-2} s^{-1}}$, which is significantly higher than the observed flux by Fermi/LAT. Thus, the stellar wind environment scenario is disfavored for GRB 180720B.}

On the other hand, for a constant density ISM environment, one has $\nu_m^{\rm IC} \propto t^{-9/4}$, $\nu_c^{\rm IC} \propto t^{-1/4}$, and $F_m^{\rm IC}\propto \tau F_m^{\rm syn}$. As $\tau=\frac{1}{3} n R\propto t^{1/4}$, we get $F_m^{\rm IC}\propto t^{1/4}$. Thus, for the ISM environment, one has
\begin{equation}
F_{\nu}= \left\{
\begin{array}{lll}
F_m^{\rm IC} \left(\frac{\nu}{\nu_m^{\rm IC}}\right)^{-\frac{p-1}{2}}\propto t^{\frac{11-9p}{8}}, \,\,\,\, \nu_m^{\rm IC}<\nu<\nu_c^{\rm IC}  \\
F_m^{\rm IC} (\frac{\nu}{\nu_c^{\rm IC}})^{-\frac{1}{2}}\propto t^{\frac{1}{8}},  \,\,\,\, \nu_c^{\rm IC}<\nu<\nu_m^{\rm IC} \\
F_m^{\rm IC} ({\nu_m^{\rm IC}})^{\frac{p-1}{2}}({\nu_c^{\rm IC}})^{\frac{1}{2}} \nu^{-\frac{p}{2}}\propto t^{\frac{10-9p}{8}}, \,\,\,\, \nu>\max(\nu_m^{\rm IC}, \nu_c^{\rm IC})
\end{array}
\right.
\end{equation}
For $p=2-2.5$, the decay slope is in the range $-1.0$ to $-1.5$. Such a slope is acceptable for GRB~180720B. For GRB~190114C, there is no sub-TeV data available yet, so we cannot distinguish between the wind model and ISM model  at present. In the following part of the paper, we will assume a constant-density  interstellar medium (ISM) for the circum-burst environment of these two GRBs.

\section{The $\gamma\gamma$ absorption and KN suppression}
Sub-TeV photons of energy $\varepsilon_\gamma$ will suffer from pair-production absorption by interacting with target photons with energy
$\varepsilon_t=\Gamma^2 (m_e c^2)^2/\varepsilon_\gamma$. For a bulk Lorentz factor of $\Gamma\sim 10-100$ at $100-10^5$ s, the energy of target photons is typically 0.1-10~keV. So the X-ray photons are the main sources to absorb sub-TeV photons. The opacity of sub-TeV photons is given by $\tau_{\gamma\gamma}=\sigma_{\gamma\gamma} (R/\Gamma) n_t$, where the number of target photons is given by $n_t=\frac{L_x}{4\pi R^2 \Gamma c  \varepsilon_t}$, where $L_x$ is the luminosity of X-ray afterglow.  Requiring $\tau_{\gamma\gamma}<1$ and using $R=4\Gamma^2 c t$, we get
\begin{equation}
\Gamma> 160 \left(\frac{L_x}{5\times10^{49}{\rm erg s^{-1}}}\right)^{1/6}\left(\frac{\varepsilon_\gamma}{\rm 1 TeV}\right)^{1/6}t_2^{-1/6}.
\end{equation}
For a  blast wave expanding in a constant-density  medium, one has $\Gamma=160 n^{-1/8}E_{54}^{1/8}t_2^{-3/8}$. Then we get
\begin{equation}
n< 1 E_{54}  \left(\frac{\varepsilon_\gamma}{\rm 1 TeV}\right)^{-4/3} \left(\frac{L_x}{5\times10^{49}{\rm erg s^{-1}}}\right)^{-4/3} t_2^{-5/3}.
\end{equation}
As the X-ray afterglows of  GRBs typically decay as  $L_x\propto t^{-\alpha}$ with $\alpha=1.2-1.4$, the constraint on the circum-burst density is insensitive to the observation time.

As the sub-TeV photons are produced by the IC process, these photons may also suffer from Klein-Nishina (KN) scattering suppression. The SSC energy output is dominated by $\gamma_m$ and $\gamma_c$ electrons, respectively, in the fast and slow cooling regime. Here $\gamma_m$ and $\gamma_c$ are, respectively, the injection break and cooling break in the electron distribution spectrum.  As pointed out by Nakar et al. (2009), the first KN break at $\nu_p^{\rm IC}=\max({\nu_m^{\rm IC}, \nu_c^{\rm IC}})$ is  very mild and  a clear steepening in the spectrum is expected to be observed at   $E_{\rm KN}=\Gamma \gamma_M m_e c^2$, where $\gamma_M=\max ({\gamma_m, \gamma_c})$.

VHE photons of GRB~180720B are detected at $t\simeq10$~hr. At such a late time, we expect $\gamma_m<\gamma_c$. Then the KN-induced break is expectedly at
\begin{equation}
E_{\rm KN}=\Gamma \gamma_c m_e c^2=0.1 {\rm TeV} (\frac{1}{1+Y_c}) \epsilon_{B,-2}^{-1}E_{54}^{-1/4} n_{-1}^{-3/4}t_{\rm 10 hr}^{-1/4},
\end{equation}
where $Y_c$ is Compton parameter for electrons with energy $\gamma_c$.
Requiring $E_{\rm KN}\ga 440{\rm GeV}$ for GRB~180720B,  we obtain
\begin{equation}
\epsilon_B\la 2\times10^{-3}(\frac{1}{1+Y_c})  E_{54}^{-1/4}n_{-1}^{-3/4} t_{\rm 10 hr}^{-1/4}(\frac{\varepsilon_\gamma}{\rm 0.44 TeV})^{-1}.
\end{equation}
Thus a low magnetic field equipartition factor is inferred for GRB~180720B. Note that the energy of the KN-induced break decreases rather slowly with time ($\propto t^{-1/4}$), which is  helpful for late-time detection of VHE photons from GRBs.

For GRB~190114C, at $t=100-1000$ s, both $\gamma_c>\gamma_m$ and $\gamma_m>\gamma_c$ are, in principle, possible.  If $\gamma_c>\gamma_m$, the above constraint is applicable.   If $\gamma_m>\gamma_c$, we have
\begin{equation}
E_{\rm KN}=\Gamma \gamma_m m_e c^2=0.3 {\rm TeV} f_p \epsilon_{e,-1}E_{54}^{1/4} n_{-1}^{-1/4}t_2^{-3/4},
\end{equation}
where $f_p=6(p-2)/(p-1)$ and $\epsilon_e$ is the fraction of shock internal energy transferred to accelerated electrons.
Requiring $E_{\rm KN}\ga 1{\rm TeV}$ for GRB 190114C, we have
\begin{equation}
\epsilon_e\ga 0.3 f_p^{-1} E_{54}^{-1/4} n_{-1}^{1/4}t_2^{3/4}(\frac{\varepsilon_\gamma}{\rm 1 TeV}).
\end{equation}
This constraint can be satisfied for sub-TeV photons detected at early times, such as those in GRB~190114C. However, for sub-TeV photons detected at late times, such as those detected at $t={\rm 10 hr}$ in GRB~180720B, this constraint is hardly satisfied.

\section{Modeling of the multi-wavelength data}

As pointed out by Ruiz-Velasco (2019), there is one striking similarity between GRB~190114C and GRB~180720B, i.e., both GRBs have very bright X-ray afterglows. This may indicate that X-ray photons serve as the synchrotron target photons for IC scatterings. The sub-TeV emission in GRB~180720B has the same level of flux as that of X-rays, indicating that the Compton parameter is close to unity.
We perform modelings of the available multi-wavelength data for GRB~180720B and GRB~190114C. The modelings are based on the numerical code that has been applied to GRB~130427A (Liu et al. 2013). In this code, a strict inverse-Compton scattering cross section has been used. The KN effect may also affect the electron distribution (Nakar et al. 2009; Wang et al. 2010), and we have calculated the electron distribution self-consistently.

{\em GRB~180720B:} The modeling results of the afterglow light curve and spectral energy distribution (SED) for GRB~180720B  are shown in  Figure 1. The synchrotron and SSC components are denoted by dotted and dashed curves respectively. The  first optical data point and the initial peak of the X-ray emission exceed the model fluxes and their emissions may be attributed to the reverse shock emission (Fraija et al. 2019b) {\footnote{Fraija et al. (2019b) attribute the X-ray peak to the synchrotron self-Compton emission from the reverse-shock region.  }}.  The break at the highest energy part of the SED corresponds to the  energy of $\gamma_c$ electrons (i.e., $\Gamma \gamma_c m_e c^2$ in the observer frame), whose value is given by Eq. 5. It can also be seen that the KN suppression starts earlier than this break, softening the spectrum from a photon index of $-(p+1)/2$ to about $-2$. Another feature is the transition from the synchrotron component to the SSC component  at about 1 GeV, above which a moderate spectral hardening is expected. However, since the GeV flux of GRB~180720B is below the sensitivity of Fermi/LAT at 10\,hrs, this transition cannot be identified in the data.

{\em GRB~190014C:} The modeling results of the afterglow light curve and spectral energy distribution (SED) for GRB~190014C are shown in  Figure 2. The optical flux of the first data point exceeds the model flux and it should be produced by the reverse shock emission (Laskar et al. 2019).  The X-ray flux at $t\la 100{\rm s}$ also exceeds the model flux and the excess flux could be attributed to the reverse shock emission as well (Laskar et al. 2019). The late-time brightening of the optical emission is not well-understood, and we speculate that  the late central engine activity might cause such a brightening (Li et al. 2012). In the LAT energy band, the model flux can explain the data at 100~s, and the early GeV emission should be attributed to the prompt emission or reverse shock emission (Fraija et al. 2019a)\footnote{During the review process of our paper, we noticed a paper  appeared on arXiv (Fraija et al. 2019c), which suggests that photons beyond the synchrotron limit in GRB 190114C can be interpreted by the SSC process. }. The plot of the SED around $t=100$~s shows  the transition from the synchrotron component to the SSC component occurs at GeV energies. Interestingly, the SSC component already contributes dominantly to the flux at energies above GeV. We derived a photon index of $-1.76\pm 0.21$ for the LAT emission during the period of 50-150 s, which might be a signature of the hard spectrum arising from the SSC emission{\footnote{A hard spectrum is also found for the high-energy emission of GRB~190114C at early times by Wang et al. (2019).}}. The sub-TeV flux expected from this SED fitting is comparable to the GeV flux, which can explain the $\ga 20 \sigma$ detection by MAGIC (Mirzoyan 2019). We note that our modeling does not suggest significant internal  $\gamma\gamma$ absorption in the source, as implied  by the gray line in Fig 2, which represents the SSC emission before considering the $\gamma\gamma$ absorption. This is different from the result in Derishev \& Piran (2019), which suggests significant pair production in the source.

{In order to see the ratio of the energies that go to SSC and synchrotron emission in our modeling,  we show, in Figure 3, the Compton $Y$ parameters for the $\gamma_c$-electrons, which produce the peaks of the SED. The $Y$ parameter taking into account the KN effect is $Y(\gamma_c)=U_{\rm syn}(\nu<\nu_{\rm KN})/U_B$, where $\nu_{\rm KN}$ is the critical
frequency of the incidence photons above which the scatterings with the $\gamma_c$-electrons enter the KN regime (Liu et al. 2013). $U_{\rm syn}$ and $U_B$ are, respectively, the energy densities of the synchrotron radiation below $\nu_{\rm KN}$ and the magnetic field. The values of $Y$  parameters for $\gamma_c$-electrons are about 0.5-1 at early times for the two GRBs, indicating that the energies radiated into the synchrotron and SSC components are comparable for these electrons.  }

{Our modeling of both GRBs gives a low magnetic equipartition factor of $\epsilon_B\sim10^{-5}-10^{-4}$. The low values of $\epsilon_B$ required for the modeling of both bursts are consistent with previous results from afterglow modeling (e.g., Barniol Duran 2014, Wang et al. 2015, Beniamini et al. 2016).  The low values of $\epsilon_B$ leads to a slow-cooling regime for the shocked electrons, i.e. $\gamma_m<\gamma_c$. } As a result, the high energy suppression due to the KN effect in the two bursts is related to $\gamma_c$-electrons. The case of $\epsilon_B < \epsilon_e$ is also the regime that significant  SSC emission is expected (Zhang \& \Meszaros 2001).

\section{Discussions and Conclusions}
It is useful to obtain the transition energy from the synchrotron component to the SSC component, as this transition energy could be identified if observation energy coverage is sufficiently wide.  This is also the critical frequency above which spectrum hardens. Assuming the transition energy is above $\nu_m^{\rm IC}$, the transition frequency ${\nu_t}$ can be obtained by
\begin{equation}
F_m (\frac{\nu_c}{\nu_m})^{-(p-1)/2}(\frac{\nu_t}{\nu_c})^{-p/2}=F_m^{\rm IC}(\frac{\nu_t}{\nu_m^{\rm IC}})^{-(p-1)/2}.
\end{equation}
Then we obtain
\begin{equation}
(\frac{\nu_t}{\nu_c})^{1/2}= \tau^{-1} \gamma_m^{-(p-1)}
\end{equation}
For $p=2.4$, we obtain
\begin{equation}
h\nu_t=15 {\rm GeV} \epsilon_{e,-1}^{-1.4}\epsilon_{B,-4}^{-1.5}E_{54}^{-0.93}n_{-1}^{-1.6}t_{2}^{-0.23}.
\end{equation}
The transition energy is sensitive to the microphysical parameters, $\epsilon_{e}$ and $\epsilon_{B}$, and also the ISM density. For some parameter space, the transition energy could be located in the Fermi/LAT energy range. In this case, one will see a hard GeV spectrum in the LAT energy range, contributed mainly by the SSC emission. The possible hard spectrum of GeV emission in GRB~190114C around $t=100$ s could be such an example. There is also tentative evidence for the presence of a hard spectral component in the GeV afterglow of GRB 130427 from  100~s up to one day after the burst (Tam et al. 2013), which  has been interpreted as arising from the SSC emission (Liu et al. 2013).

The two bursts that have sub-TeV photons share some common features: 1) both have low redshifts, which is useful to avoid the EBL absorption; 2) both are strong bursts with high fluence; 3) the circum-burst medium of both bursts is likely a uniform-density ISM, rather than a stratified stellar wind. These properties may explain the rare detection of VHE photons so far. On the other hand, the detection of high-significance  sub-TeV emission at 100--1000 s from GRB~190114C and the late-time detection (at 10\,hr) from GRB~180720B are inspiring for ground-based VHE observations.  These detections demonstrate that the IC component of the afterglow emission is as strong as the synchrotron component. The detection at late times also implies that the KN suppression and $\gamma\gamma$ absorption to VHE emission does not increase with time. This is important for long-term VHE observations of GRBs.

\section*{Acknowledgments}
We would like to thank T. Piran, E. Derishev, and N. Fraija for useful discussions. X.Y.W. is supported by the National Key R \& D program of China under the grant 2018YFA0404203 and the NSFC  grants
11625312 and 11851304.

\clearpage
\begin{figure}
\centering
\includegraphics[scale=0.5]{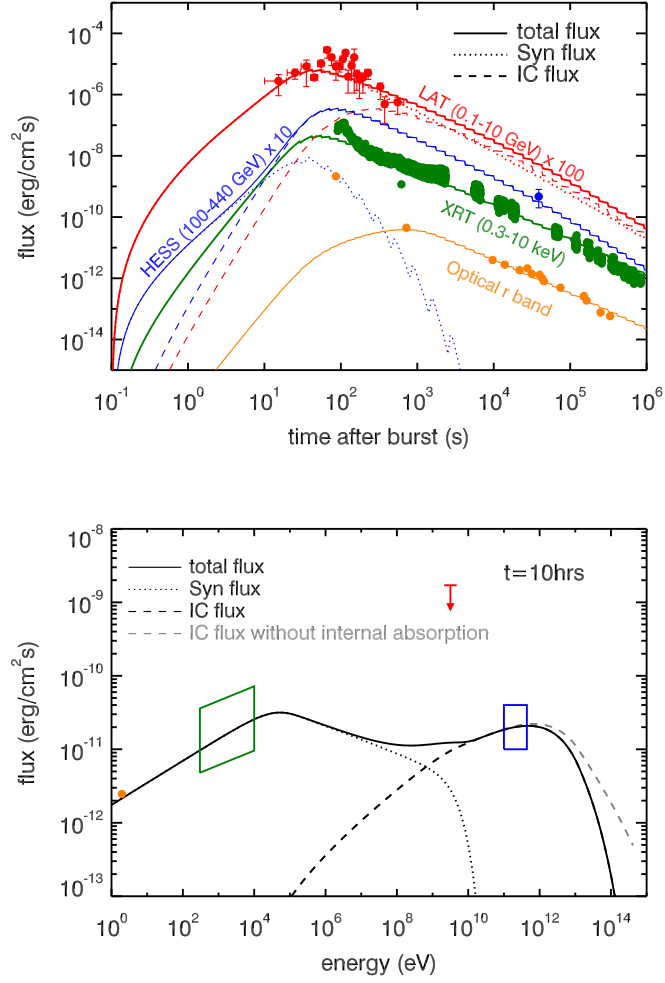}
\caption{{\bf Upper panel}: Modeling of the broad-band afterglow light curves of GRB~180720B. The LAT, HESS and optical data are taken from  Ruiz-Velasco (2019), and the XRT data is retrieved from Swift-XRT GRB light-curve repository (http://www.swift.ac.uk/xrt\_curves).
The dotted curves and dashed curves represent the synchrotron component and SSC component, respectively. For visibility, the data and theoretical fluxes in the LAT band and in HESS band are multiplied by 100 and 10 respectively. The  theoretical curve of the optical afterglow emission has been corrected to account for the extinction by the host galaxy (assuming $A_V=0.8{\rm mag}$). {\bf Bottom panel}: Modeling of the afterglow SED of GRB~180720B at $t=10{\rm hr}$. The green and blue boxes represent the X-ray data and HESS data, respectively. The upper limit is from the non-detection of Fermi/LAT. The gray dashed curve represents the SSC emission before considering the $\gamma\gamma$ absorption in the source.   The parameters used in the fitting are: $E=10^{54} {\rm erg}$, $n=0.1{\rm cm^{-3}}$, $\epsilon_e=0.1$, $\epsilon_B=10^{-4}$, $\Gamma_0=300$ and $p=2.4$. }
\end{figure}

\begin{figure}
\centering
\includegraphics[scale=0.5]{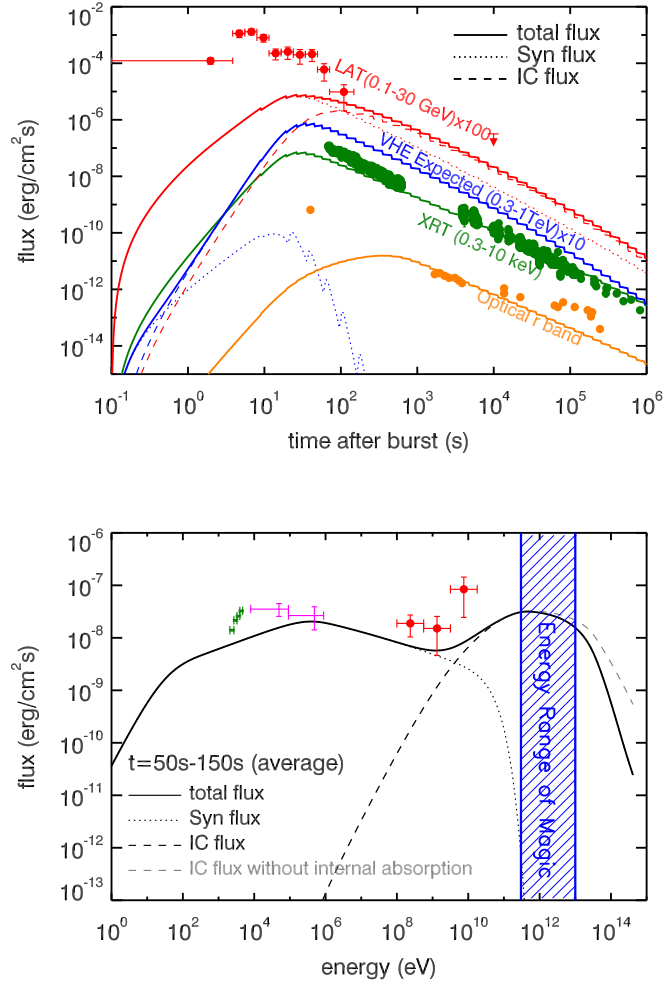}
\caption{{\bf Upper panel}: Modeling of the broad-band afterglow light curves of GRB~190114C. The optical data are taken from  Laskar et al. (2019), the X-ray data are retrieved from Swift-XRT GRB light-curve repository, and the LAT  data are reduced by ourself. The dotted curves and dashed curves represent the synchrotron component and SSC component, respectively. The  theoretical curve of the optical afterglow emission has been corrected to account for the extinction by the host galaxy (assuming $A_V=2.2{\rm mag}$). For visibility, the data and model theoretical fluxes in the LAT band and in VHE band are multiplied by 100 and 10, respectively. {\bf Bottom panel}: Modeling of the afterglow SED of GRB~190114C during the period of 50-150 s. The green dots represent the X-ray data. The purple data represent the GBM data, which are reduced by ourself. The red circles represent the GeV data of Fermi/LAT. The error-bars of of these data correspond to $1\sigma$ confidence level. The blue hatched region is the energy range of the Magic telescope. The gray dashed curve represents the SSC emission before considering the $\gamma\gamma$ absorption in the source. The parameters used in the fitting are: $E=6\times 10^{53} {\rm erg}$, $n=0.3{\rm cm^{-3}}$, $\epsilon_e=0.07$, $\epsilon_B=4\times 10^{-5}$, $\Gamma_0=300$  and $p=2.5$. }
\end{figure}

\begin{figure}
\centering
\includegraphics[scale=0.5]{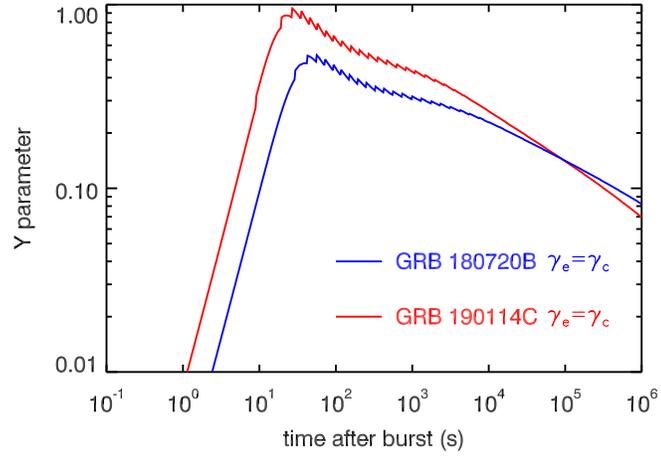}
\caption{Compton $Y$ parameters as a function of time for the electrons with a Lorentz factor of $\gamma_c$. }
\end{figure}

\end{document}